# Coulomb blockade-like transport and enhanced memory in organic transistors embedded with sub-nm Pt nanoparticles for neuromorphic computing

Arash Ghobadi[†], Thomas B. Kallaos[†], Abhi Abhijeet[†], Stephen C. Klue[†], Joseph C. Mathai[#], Carsten A. Ullrich[†], Shubhra Gangopadhyay[#*], and Suchismita Guha[†‡*]

[†]*Department of Physics and Astronomy, University of Missouri, Columbia, MO 65211*

[‡]*MU Materials Science and Engineering Institute, University of Missouri, Columbia, MO 65211*

[#]*Department of Electrical Engineering and Computer Science, University of Missouri, Columbia, MO 65211*

[*] *Corresponding Author E-mail:* gangopadhyays@missouri.edu; guhas@missouri.edu

**Abstract**

Organic transistors are playing an increasingly important role for neuromorphic applications. However, devices that rely solely on ferroelectric switching or on interface traps for their multi-conductance states exhibit limited memory windows. Here, we introduce an ultrathin oxide layer with a uniform distribution of sub-nm platinum nanoparticles (PtNPs) at the interface of a polymer semiconducting and a ferroelectric dielectric in a thin film transistor architecture. The interfacial stack, $Al_2O_3$/PtNP/$Al_2O_3$, provides a viable route for localized charge trapping and de-trapping in a region where it can most effectively influence the channel conductance. The organic transistors display a large memory window (> 20 V) in their current-voltage characteristics. The sub-nm PtNPs give rise to features that are consistent with room temperature Coulomb blockade-like transport, supporting discrete and well-separated levels within the memory window. The devices support multimodal programming using electrical and optical stimuli with both long-term plasticity and enhanced short-term plasticity (STP) phenomena. These results open new directions for implementing STP in the development of neuromorphic computing.

The next-generation memory chips increasingly need on-chip elements that can store analog states, offer a wide and well-controlled memory window, and operate with very low power.[1-3] Equally important in this direction of enhancing memory operation is to reduce the cost and ease of fabrication of electronic devices. Organic transistors with solution processed semiconducting layers provide a pathway towards low-cost electronics. However, due to the inherent disordered transport in the semiconducting layer along with the influence of the dielectric layer which may alter the density of states, achieving a stable memory window with high repeatability using organic transistors may be challenging. Hence, a robust interface layer between the semiconductor and the dielectric that can trap and de-trap charge carriers is essential for potential memory and neuromorphic applications.

Nanoparticles (NPs) at the interface of the dielectric and the semiconductor layer in a two-terminal or a three-terminal device have been used for enhancing the memory window.[4-7] A well-controlled layer of NPs further allows for exploiting quantum mechanical effects, which is perhaps even more relevant after the 2025 Nobel Prize in Physics. The single electron transistor (SET)[8] has been the hallmark of the Coulomb blockade effect,[9,10] where the tunneling current across the source-drain junction diminishes due to a strong Coulomb repulsion between charges, increasing global energy. By tuning the gate voltage to a charge equivalent that is half the electronic charge to the gate capacitance, the tunneling current again increases. Observing Coulomb blockade at room temperature is challenging mainly due to the extremely small dimensions of the source-drain contacts that are required in an SET[11,12] or due to the requirement of a well-controlled active layer in a transistor in the presence of metal NPs for isolating charge carriers in the device.[13]

As the field of neuromorphic devices continues to expand, dual mode programming that integrates both electrical and optical stimuli has attracted significant attention.[14-16] This approach broadens device functionality and enables biologically inspired information processing capabilities. Organic semiconductors with tunable band gap energies and high absorption coefficients are an excellent choice for photonic synapses. The ability to unify electrical and optical programmability in one compact device expands the functional scope of organic neuromorphic hardware and brings it closer to practical and multifunctional systems. Neuromorphic systems emulate key functional aspects of neural computation, including distributed parallelism, adaptive state evolution, and local memory dependent signal transformation.[17] The objective of such systems is to identify device and circuit mechanisms that can implement neural network relevant operations more efficiently than conventional digital approaches for targeted tasks. In practice, this often translates to developing electronic devices whose conductance can be tuned, retained, and updated in a controlled manner so that they can serve as hardware analogs of synaptic weights. Current thrust in neuromorphic devices relies on nanomaterials such as low-dimensional semiconductors, nanowires, and metal nanoparticles. Such systems closely mimic biological systems with improved characteristics to replicate neural networks.[18]

Synaptic plasticity forms the basis of learning and memory as it provides a mechanism by which transient neural activity can result in persistent changes in a circuit function, shaping subsequent computation and behavior.[19-21] The synaptic strength, a measure of synaptic plasticity, can be both long-term and short-term. Organic ferroelectric transistors employing organic semiconductors as the active layer and polymeric ferroelectrics as the dielectric layer typically display long-term plasticity (LTP) effects,[22,23] mainly due to the slow polarization dynamics with conductance states lasting longer than short-term plasticity (STP) effects. STP, where the synaptic strength changes in millisecond to second timescales, correlates well with computations related to speech recognition and working memory.[24,25] Unlike LTP characteristics, which have been heavily exploited in synaptic devices, the implementation of STP characteristics at the device level has been limited.[26] In biological nervous systems, paired-pulse facilitation (PPF)[27] exemplifies STP that demonstrates the capacity of enhancing the postsynaptic response of the second stimulus compared to the first in the presence of two closely spaced presynaptic stimuli. PPF plays an important role in signal transmission and information processing because repeated stimulation can transiently enhance synaptic weight, thereby promoting stronger and more reliable propagation of neural signals over short time scales.[28,29]

In this work, we demonstrate room temperature Coulomb blockade-like transport along with enhanced STP characteristics in organic transistors. By using bottom-gate, top-contact organic transistors, sub-nm PtNPs are incorporated as tunable charge storage sites at the interface within two ultrathin aluminum oxide layers between the organic semiconducting (DPP-DTT) and ferroelectric dielectric (PVDF-HFP) layers. The PtNPs, deposited by controlled magnetron sputtering, are sandwiched between aluminum oxide thin films: $Al_2O_3$/PtNPs/$Al_2O_3$ creates a localized charge trapping and de-trapping region at the semiconductor/dielectric interface, which directly couples to the channel region. The resulting transistors exhibit an enhanced and tunable memory window, robust multi-level conductance states, improved retention, and highly repeatable switching under electrical and optical pulsed operation. Due to the sub-nm size of the PtNPs, the devices show features that are consistent with quantum phenomena at room temperature, such as Coulomb blockade-like transport with distinct plateaus in the current-voltage characteristics, which further sharpen the memory behavior and sustain discrete, well-separated levels. At the same time, this platform supports multimodal programming using both electrical and optical stimuli for modulating the multi-conductance states with LTP and enhanced STP phenomena. The PtNP-engineered devices can be programmed electrically through gate voltage pulse schemes and can also be modulated optically through pulsed illumination, enabling a single hardware platform to combine non-volatile memory, synaptic plasticity, and photoresponsive functionality.

## Device architecture and attributes of sub-nm PtNPs

Organic thin film transistors with multi-conductance states originating from a ferroelectric dielectric layer, trapping states, or due to ionic contribution mimic biological synapses (Fig. 1a), where the gate terminal emulates a presynaptic neuron and the source-drain channel is similar to a postsynaptic neuron. PPF, a characteristic of STP, relies on the concentration of $Ca^{2+}$ controlled by the voltage-gated channel. As shown in Fig. 1a, the PPF is triggered by an action potential at the presynaptic terminal. The organic transistors with DPP-DTT as the semiconductor and PVDF-HFP as the dielectric layer were fabricated as a bottom-gate, top contact geometry with Al as the gate electrode and Au as source/drain electrodes (Fig. 1b). A 12 nm $Al_2O_3$ film was grown using atomic layer deposition (ALD) on top of the spin-casted PVDF-HFP layer with optimized growth parameters.[30] PtNPs were deposited on the $Al_2O_3$ layer using a magnetron sputtering system with a tilted target (23.8°) configuration for two different deposition times of 10 s and 20 s. The optimization of PtNPs was achieved in prior works.[31,32] Figure 1c shows a high-resolution transmission electron microscope (HRTEM) image of PtNPs deposited for 10 s. The average particle size is $0.66 \pm 0.26$ nm with an areal particle density of ~$10^{12}$ $cm^{-2}$. The 20 s deposition time yields slightly larger size particles of $1.30 \pm 0.43$ nm (Supplementary Fig. 1). The smaller sized PtNPs (~ 0.7 nm) are less crystalline compared with the 1.30 nm particle size (Supplementary Fig. 2). The PtNPs were capped with a 2 or a 3 nm ALD grown $Al_2O_3$ film, which acts as the tunneling layer. 50 nm thickness of DPP-DTT was spincoated on top of the $Al_2O_3$/PtNPs/$Al_2O_3$ stack, followed by evaporation of the source/drain contacts. Transistors without PtNPs but with a 12 nm ALD grown $Al_2O_3$ interface between the DPP-DTT and the PVDF-HFP layer served as control devices. The channel length varied between 50 µm – 125 µm. All transistors show *p*-type transport.

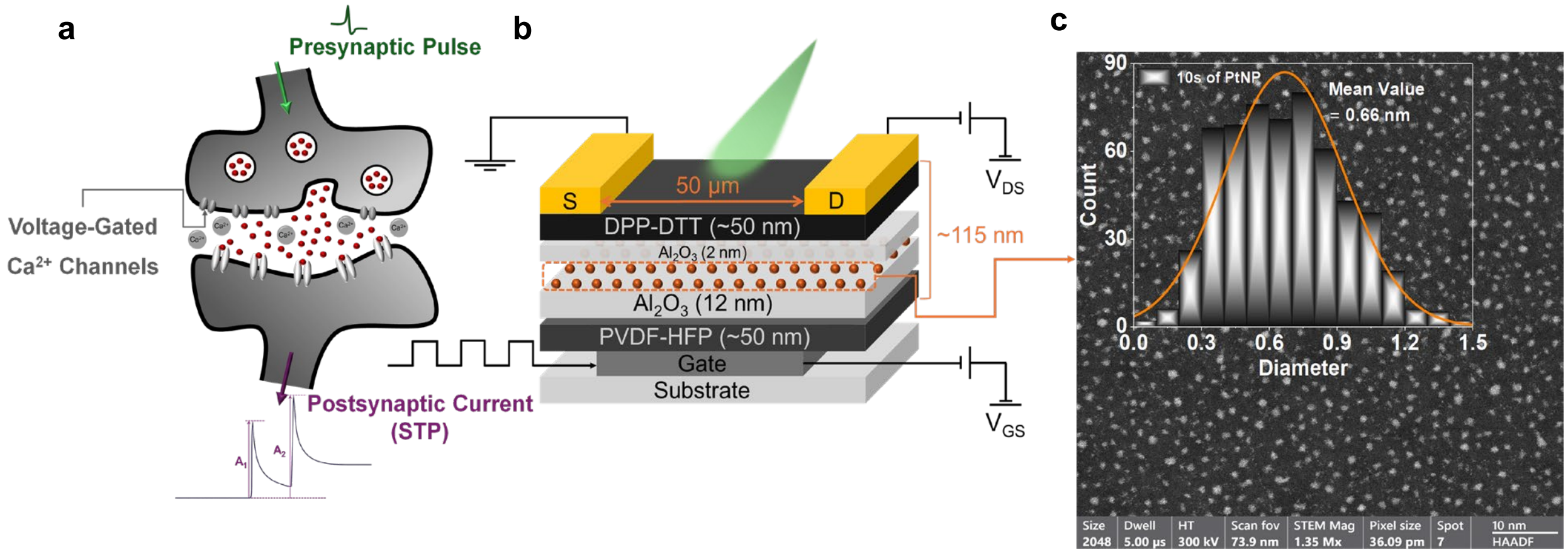

**Fig. 1| Organic transistors with PtNPs mimic short-term plasticity. a**, Schematic of a biological neuron with the release of neurotransmitters mediated by calcium ions that results in short-term plasticity (STP). **b**, Device architecture of organic transistor with DPP-DTT as the semiconducting layer and PVDF-HFP as the gate dielectric. ALD grown $Al_2O_3$ (12 nm thickness) sputtered with PtNPs along a 2 nm ALD grown $Al_2O_3$ tunneling layer is placed between the

semiconducting and the dielectric layer. The presynaptic signal, either as a pulsed voltage at the gate or a modulated illumination at the channel area, results in postsynaptic current. **c**, HRTEM image of PtNPs deposited for 10 s cycle. The inset shows a histogram of the particle size distribution.

## Enhanced memory window

Figure 2a-c compares the transfer characteristics from a control device (without PtNPs) and two other PtNP embedded devices with 2 nm and 3 nm $Al_2O_3$ tunneling layers. The transfer hysteresis loops, where the gate-source voltage ($V_{GS}$) was varied from 2 V to 20 V, are shown below the transfer curves for each of the three devices. The transistor parameters were extracted using the Mott-Schottky analysis (Supplementary Table I). A clear difference between the control device and the PtNP embedded device is in the sign of the threshold voltage ($V_{th}$). Typically, $V_{th}$ is negative for *p*-type transistors. From the transfer sweeps of the PtNP embedded 2 nm and 3 nm devices (Fig. 2b, c), $V_{th}$ is found to be +4.62 V and +6.62 V, respectively. The positive values of $V_{th}$ indicate the PtNPs to be negatively charged, effectively making the channel be initially positively charged. Additionally, the gate leakage current ($I_G$) is similar for devices with 0.7 nm PtNP embedded transistors and the control device (Supplementary, Fig. 3).

Current-voltage (*I-V*) hysteresis measurements were achieved by keeping the drain-source voltage ($V_{DS}$) constant at -7 V and by sweeping the gate voltage ($V_{GS}$) between positive and negative values, starting at 2 V → -2 V to 20 V → -20 V (Fig. 2d-f). As expected, the control device shows no hysteresis whereas the PtNP embedded devices display a large hysteresis, characteristic of a non-volatile memory (NVM). As the gate voltage is swept from positive to negative values, a build-up of minority and majority carriers occurs at the semiconducting-$Al_2O_3$ tunneling layer. Starting with positive $V_{GS}$, the PtNPs are negatively charged (Fig. 2g), and as $V_{th}$ is reached, the flat-band condition is introduced (Fig. 2h). As $V_{GS}$ is swept to more negative voltages, accumulation of holes occurs in the channel while the PtNPs retain their charge, which is commensurate with the writing process of the NVM. As the channel gets depleted during the reverse sweep, the PtNPs release their charges (Fig. 2i) representing the erasing process (discussed in more detail in the next Section). Since the $Al_2O_3$ tunneling layer is less than 3 nm, the charge carriers can directly tunnel in and out of the PtNPs.[32]

The memory window is more robust with the 3 nm $Al_2O_3$ tunneling layer compared with the 2 nm oxide layer, which originates from a higher retention of charges in the PtNPs for the thicker oxide layer. It is further seen that by reducing $V_{DS}$ below -1 V, the NVM window can be enhanced although, as expected, the overall accumulation current decreases by a few orders of magnitude (Supplementary, Fig. 4). The large memory window in these devices is particularly favorable for modulating the multi-conductance states with optical stimuli, enhancing STP effects, as shown later. Although the devices were stored in a glove box, all measurements were conducted in air; the *I-V* characteristics remain almost unchanged for more than 5 months.

A distinctive feature of all *I-V* sweeps is the presence of plateaus with spikes in the current at regular intervals during the reverse sweep (from accumulation to depletion of the channel). These spikes occur in two or three steps depending on the sweep rate. The plateaus can be explained based on the ionization potential (IP) and the electron affinity (EA) of the PtNP clusters, whereas the evenly spaced current spikes correspond to a channel screening due to the oscillation of the net electric field as the charge of the PtNPs change. We will discuss these features in detail in the next section.

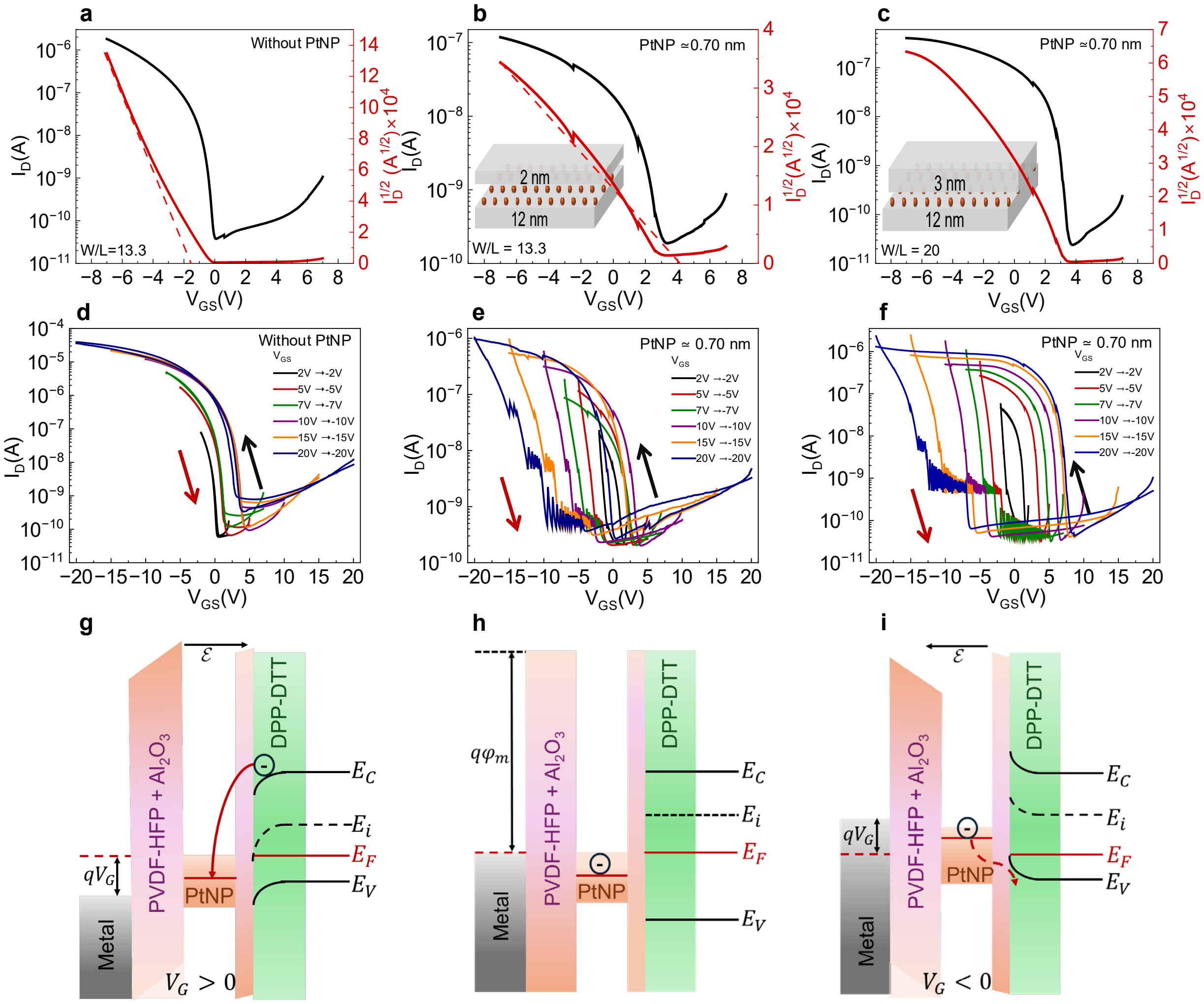

**Fig. 2| Transfer sweeps and memory window. a**, Transfer characteristics (black curve) of the controlled DPP-DTT transistor without any PtNPs. The red curve shows the square-root of the transfer plot. The intercept of the dotted red line on the x-axis denotes $V_{th}$. **b**, Transfer characteristics (black curve) of the PtNP embedded DPP-DTT transistor with 2 nm $Al_2O_3$ tunneling layer. The red curve shows the square-root of the transfer plot. The intercept of the dotted green line on the x-axis denotes $V_{th}$. **c**, Transfer characteristics (black curve) of the PtNP embedded DPP-DTT transistor with 3 nm $Al_2O_3$ tunneling layer. The red curve shows the square-root of the transfer plot. $V_{DS}$ was at -7 V for all three transistors. **d**, Transfer characteristics by sweeping $V_{GS}$ from positive to negative values and then back to the positive value for 5 different loops, starting from 2 V to 20 V in the control device. **e**, Transfer characteristics sweep by changing $V_{GS}$ from positive to negative values and then back to the positive value in PtNP embedded device with 2 nm $Al_2O_3$ as the tunneling layer. **f**, Transfer characteristics sweep by changing $V_{GS}$ from positive to negative values and then back to the positive value in PtNP embedded device with 3 nm $Al_2O_3$ as the tunneling layer. In both PtNP embedded transistors, a large hysteresis is seen in the I-V curves. **g**, Schematic band diagram of a PtNP embedded

device with positive $V_G$. Here the PtNP gains a negative charge. **h**, Schematic band diagram of a PtNP embedded device at $V_{th}$, where the PtNP retains its negative charge. **i**, Schematic band diagram of a PtNP embedded device with negative $V_G$, representing the reverse sweep when the channel current rapidly falls during the erasing stage. The external electric field ($\mathcal{E}$) is depicted by the black arrow on top.

## Coulomb blockade-like transport

Figure 3a shows the reverse $I$-$V$ sweep for a PtNP (0.7 nm) embedded transistor with 2 nm $Al_2O_3$ tunneling layer. The two striking features here are the presence of the plateaus shown by the blue rectangles and the regular spikes in the channel current. The steps occur when the current changes by an order of magnitude (notice the logarithmic scale). The zoomed-in regions show the spacing between the current spikes between 0.3 V-0.15 V. These spikes are not seen in the control device, indicating that they originate from the PtNPs. A comparison of the 0.7 nm and 1.30 nm embedded PtNP devices (for identical channel lengths), as shown in Supplementary Fig. 5, presents a clear distinction in terms of the width of the plateaus as well as the separation between the individual current spikes. Although it is not an SET, the plateaus observed in the transport measurements are similar to a Coulomb-blockade behavior, arising from the charging and discharging of PtNPs.

Our results may be understood based on a prior density functional theory (DFT) calculation of PtNP clusters, where detailed structure, energetics, and the electronic structure of three cluster sizes of $Pt_n$ ($n$ = 13, 38, and 55) were investigated.[33] These calculations explicitly consider optimized structures, yielding the energetics of the neutral clusters and their singly positively and negatively charged ionic counterparts. Specifically, $Pt_{13}$ with an average diameter of 0.8 nm, $Pt_{38}$ with an average diameter of 1.1 nm, and $Pt_{55}$ with an average diameter of 1.3 nm are within the experimental PtNP sizes for the 10 s and 20 s deposition times. The EA, given by the change in energy: $E(N+1) - E(N)$, where $N$ is the number of electrons, and the IP, which is the change in energy: $E(N) - E(N-1)$, are estimated to be ~ 3 eV and 7 eV for the $Pt_{13}$ (octahedral) cluster, respectively. These energies, together with those of $Pt_{38}$ and $Pt_{55}$, fit very well within the spherical droplet model,[34] where theory predicts a linear relationship with the inverse radius of the PtNP clusters. Furthermore, a 20 % change in the cluster size has almost no impact on the EA and the IP energies.

Upon comparing our experimental results with the above theory, we note that although there is a size distribution of the PtNPs, the IP and EA energies should be relatively constant. We compare our experiments with $Pt_{13}$ as a typical representative of the PtNPs. The two plateaus shown in Fig. 3a agree well with the charge state of the PtNP clusters. The width of the first plateau, which approximately starts at – 10 V, is ~ 2.3 V, similar to the EA energy of the $Pt_{13}$ cluster (Fig. 3b). The second plateau at lower voltages is ~ 5.7 V wide and agrees with the IP energy of the $Pt_{13}$ cluster. It should be noted that the measured EA and IP energy values are smaller than in DFT, which is not surprising since the calculations are in the gas phase whereas the actual clusters are embedded in a dielectric medium. We thus conclude that the sudden decrease in current at – 8 V reflects a change in the charge state from $Pt_{13}^{-} \rightarrow Pt_{13}^{0}$, whereas the decrease in current at -2.4 V reflects a change from $Pt_{13}^{0} \rightarrow Pt_{13}^{+1}$. These changes in the PtNP charge states are schematically shown in Fig. 3c. Furthermore, it is instructive to compare the width of the plateaus for the 0.7 nm and 1.3 nm PtNP devices (Supplementary, Fig. 5). Similar to the prediction by DFT calculations, where the EA energy of $Pt_{58}$ is found to be ~ 30 % higher compared with $Pt_{13}$,[33] we observe a larger width of the first plateau for the 1.3 nm PtNP embedded device compared with the 0.7 nm device. A third plateau region at higher voltages, often seen in larger channel lengths, most likely reflect a change from the Pt doubly ionized negative state to $Pt^{-}$.

To understand the origin of the individual spikes, one can assume the PtNPs to be a sheet of charge (Fig. 3d), acting like a floating gate which screens the applied field of the channel. Using Gauss' law for a sheet of charge embedded in a dielectric environment, the electric field of the sheet of PtNPs is given by $\mathcal{E} = \frac{\sigma}{2\epsilon_0\epsilon}$, where σ is the areal charge density, $\epsilon_0$ is the permittivity of free space, and $\epsilon = 10$ (for $Al_2O_3$). Assuming 1 electron per cluster and with an areal density of PtNPs as $10^{12}/cm^2$, $\sigma = 1.602 \times 10^{-3}$ $Cm^{-2}$, and $\mathcal{E} = 9.05 \times 10^{6}$ V/m. With the dielectric thickness below the PtNP sheet as approximately 65 nm (PVDF-HFP+$Al_2O_3$), the voltage of the sheet is ~ 0.6 V, similar to the spacing between the spikes. As such, the electrons from the PtNPs as they tunnel back and forth do not contribute to the steady state channel current, but the channel experiences an oscillating screened voltage, which is reflected in the channel current. As the energy is close to EA or IP at the plateaus, the electron from the PtNPs tunnel in and out, modifying the net voltage. Next, we use a phenomenological model to better understand the oscillatory nature of the current.

Within each oscillatory region, the current repeatedly rises in rapid spikes, modeled as $I_{rise}$, and decays back to a floor current $I_{floor}$, modeled as $I_{decay}$.

$$I_{rise} = I_{leak} + I_n\left(1 - e^{-(V-V_n)/\tau_{rise}}\right) \quad (1)$$

$$I_{decay} = I_{leak} + I_n e^{-(V-V_n)/\tau_n} \quad (2)$$

Underlying the oscillatory behavior is a gate voltage independent leakage current $I_{leak}$, setting the baseline upon which all spikes are superimposed. Each spike and its subsequent decay are indexed by an integer $n$; a new spike initiates when the drain current returns to $I_{floor}$. The gate voltage at which the current returns to $I_{floor}$ defines the starting voltage $V_n$ of the next spike, such that each cycle begins exactly where the previous decay ends. The amplitude of each successive spike attenuates by a fixed fraction $\alpha$, such that $I_{n+1} = I_n(1-\alpha)$. The rise constant $\tau_{rise}$ is held constant across all spikes. The voltage decay constant: $\tau_n = \tau_{steady} + \left(\tau_{initial} - \tau_{steady}\right)e^{-\lambda n}$ relaxes as a function of $n$ from an initial value $\tau_{initial}$ toward a steady-state value $\tau_{steady}$, with the transition voltage rate governed by $\lambda$. For all devices the model achieves an RMS residual below 1%.

Figure 3e shows the experimental data from a 0.7 nm PtNP embedded transistor (with a 2 nm $Al_2O_3$ tunneling layer) along with a fit to eqns. 1 and 2. Here, $\tau_{initial}$ is 0.203 V, $\tau_{steady}$ is 0.086 V, and $\lambda$ is 0.327 V. Across all PtNP embedded devices, the fits to the oscillatory part of the current yields a similar value of $\lambda$, mimicking the average spacing of the current spikes. The presence of such regular spikes that can be described so well by the above model suggests a remarkable uniformity and collectivity in the tunneling behavior of the PtNPs.

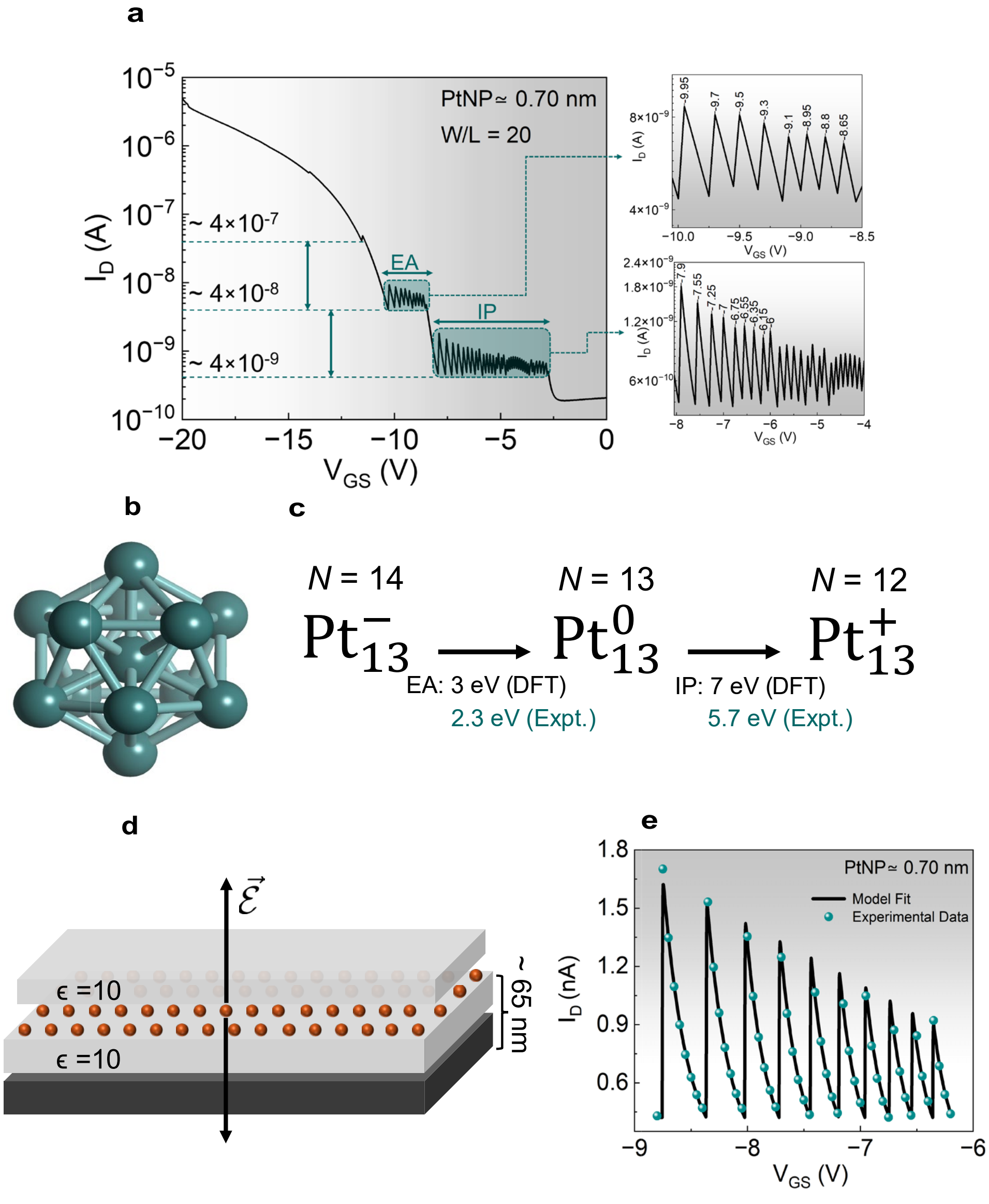

**Fig. 3| Coulomb-blockade-like transport. a**, Reverse-sweep transfer characteristics of a PtNP embedded transistor with 2 nm $Al_2O_3$ tunneling layer. The two plateaus with oscillatory current behavior observed are zoomed in on the right. **b**, A $Pt_{13}$ cluster is plotted in VESTA.[35] **c**, A comparison of the EA and IP energies of a $Pt_{13}$ cluster estimated from DFT[33] and experimental data. **d**, The PtNPs between the two $Al_2O_3$ layers (light gray slabs) on top of the PVDF-HFP layer (dark gray slab) act as a sheet of charge resulting in an electric field which screens the applied field. **e**, The

oscillatory current behavior from 0.7 nm PtNP embedded transistor. The green spheres represent the experimental data and the bold black lines are a fit to the model described by eqns. 1 and 2.

## Dual mode programming

Building on the memory behavior described above, the 0.7 nm PtNP embedded transistors were examined under both optical and electrical stimuli to evaluate their dual programmability for memory and neuromorphic applications. The PtNP layer introduces a localized interfacial charge-storage element at the semiconductor and dielectric interface, enabling optical writing and electrical erasing within the same transistor. This dual-stimulus operation is important since it allows the device conductance state to be programmed and erased through different external inputs, relevant for optoelectronic memory and neuromorphic hardware.

Photogenerated carriers are produced in the semiconducting layer upon optical illumination, while the PtNPs and other defect states act as localized trapping sites that can retain a fraction of these carriers, tunneling through the $Al_2O_3$ barrier. The transfer sweeps with and without light (532 nm) for a PtNP embedded transistor with 2 nm $Al_2O_3$ as the tunneling layer are compared in Supplementary, Fig. 6. The off current in the presence of light is higher due to photogenerated carriers, and $V_{Th}$ shifts from 0.8 V under dark conditions to 11.6 V in the presence of the 532 nm light. The stored charge modifies the local electrostatic environment near the channel and increases the drain current, corresponding to an optical potentiation or writing. In contrast, electrical gate pulses can promote charge de-trapping or redistribution of charges at the semiconductor-dielectric interface, reducing the stored charge and lowering the drain current. Hence, the $Al_2O_3$/PtNPs/$Al_2O_3$ interfacial stack provides a common charge-storage platform, supporting both optical programming and electrical erasing.

A train of optical and electrical pulses was applied to the device by biasing the gate and keeping $V_{DS}$ constant at −7 V for evaluating LTP properties. The optical writing process was assessed under both green and blue illumination by switching on and off at 1s intervals. Similarly, the electrical erasing was carried out by applying gate pulses ($V_{GS}$ = +2V) for 1s on and off intervals. Figure 4 displays both LTP and STP characteristics from a 0.7 nm PtNP embedded transistor (with a 3 nm $Al_2O_3$ tunneling layer). The currents shown here were measured by a lock-in amplifier technique and are mainly displayed as representative plots. For pattern recognition accuracies and neural network training, the currents were measured using a sensitive source meter (details are provided in Supplementary Information).

The optical writing and electrical erasing under 532 nm illumination with a power density of 0.03 W/cm² is shown in Fig. 4a, where the drain current progressively increases during optical stimulation and decreases during the subsequent electrical erasing sequence. The zoomed-in region of a few pulses indicates that the conductance states are updated incrementally rather than through a single abrupt transition. A similar optical writing and electrical erasing response was also observed under 405 nm illumination with a power density of 0.008 W/cm², as shown in Fig. 4b. These results confirm that the PtNP embedded transistors exhibit dual programmability under different optical wavelengths while maintaining electrical erase functionality.

The experimentally measured synaptic responses were further benchmarked using NeuroSim+[36] to evaluate their potential for image recognition. The main input parameters are determined by the maximum ($G_{max}$) and minimum conductance ($G_{min}$), which are governed by nonlinear equations during potentiation ($P$) and depression ($D$), and are given by: $G_P = B[1 - \exp(-p/A)] + G_{min}$ and $G_D = -B[1 - \exp(p-1)\,/A] + G_{max}$. Here, $p$ represents the number of pulses; $B = (G_{max} - G_{min})/(1 - \exp(-1/A))$. The conductance and pulse numbers were normalized to 1 and the normalized nonlinearity values: $A$ and $B$ were obtained by fits with the above nonlinear equations during the potentiation and depression cycles. The actual nonlinear parameters during potentiation and depression are inversely related to the $A$ and $B$ values,[36] and are depicted by $\beta_p$ and $\beta_d$ as shown in Supplementary, Fig. 7.

Using the MNIST (Modified National Institute of Standards and Technology) handwritten digit data set, a two-layer multilayer perceptron neural network was employed for online learning and offline classification. As shown schematically in Fig. 4c, the network consists of 400 input neurons by considering 20 × 20 pixel images, 100 hidden-layer neurons, and 10 output neurons. The output neurons represent the digit classes from 0 to 9. The network was trained using 10,000 images during each epoch.

The simulated image recognition accuracies obtained from the experimentally extracted conductance updates are shown in Fig. 4d. After 125 training epochs, the device response based on the 532 illumination and electrical depression achieved an accuracy of approximately 83%, whereas the response based on the 405 nm optical writing reached approximately 36%. This difference is mainly attributed to the large dynamic range obtained under 532 nm optical programming under our testing conditions, which provides the overall weight update during the neural network training. We note that although the overall current was lower under the 405 nm illumination due to an order of magnitude lower input power density, it is the dynamic range ($G_{max}/G_{min}$) between the first pulse and the saturation pulse that plays a role, which scales quite uniformly for different incident power densities. Moreover, the relationship between the $\beta_p$ values and the dynamic range affects the overall accuracies.[22] Device-to-device variation in pattern recognition accuracies with the 532 nm are shown in Supplementary, Fig. 8. The ideal device shown in Fig. 4 d is a simulation for a ferroelectric transistor with no nonlinearity in its conductance.

Unlike long-term potentiation with optical pulses, electrical pulses do not induce such an effect. The bi-directional tunneling of carriers through the ultrathin $Al_2O_3$ tunneling layer upon the application of a bias may result in charge relaxation at the highly smooth 12 nm $Al_2O_3$ barrier and PVDF-HFP, preventing potentiation. Additionally, the slow traps at the interface could be inaccessible for carriers with the application of electrical potentiation pulses. Strategies for enhancing defect states at the tunneling layer, including increasing the thickness of the tunneling layer, without decreasing the memory window, in the future could be potential avenues for fully electrical synaptic devices.

In addition to the optical writing and electrical erasing, the PtNP embedded transistors exhibit STP. As a representative short-term synaptic function, the PPF was examined under optical excitation. Paired optical pulses (532 nm, 0.03 W/cm$^2$) were applied with 1 s on/off intervals at $V_{GS}$ = +1 V, while $V_{DS}$ was held constant at −7 V. Fig. 4e shows the current-time response under the paired-pulse protocol. The zoomed-in image on the right illustrates the definition of the first and second post-synaptic response amplitudes, $A_1$ and $A_2$, relative to the baseline current.

The PPF index was calculated as $(A_2/A_1) \times 100\%$, where a larger value indicates stronger facilitation. In this device, the PPF index reached as high as 210%, meaning that the second post-synaptic response was approximately 2.1 times larger than the first. In biological synapses, a similar enhancement is associated with residual $Ca^{2+}$ remaining after the first spike, which increases the neurotransmitter release probability during the second spike. In the present PtNP embedded transistor, the analogous behavior is attributed to residual photogenerated carriers and incompletely relaxed trapped charge that remain after the first optical pulse, thereby amplifying the response to the second optical stimulus.

The transient nature of the facilitation indicates that the PtNP-based interfacial stack supports short-term synaptic memory in addition to long-term conductance modulation. The decay in the PPF signal was fitted using a double-exponential model,[26] which reflects the coexistence of fast and slow relaxation processes. The extracted time constants were $\tau_1 = 0.71 \pm 0.27$ s and $\tau_2 = 2.34 \pm 0.89$ s, indicating that the facilitation dynamics occur on sub-second to few-second time scales. These values are consistent with short-term biological plasticity,[37] observed in other organic transistors as well,[14,38] and suggest that multiple trap-assisted relaxation pathways contribute to the transient synaptic response.

Overall, PtNP embedded organic transistors demonstrate optical writing, electrical erasing, wavelength-dependent LTP and STP within the same device platform. The coexistence of long-term programmability and short-term facilitation highlights the role of the PtNP-based interfacial charge-storage layer in enabling multifunctional optoelectronic synaptic behavior. We further compare the performance of the PtNP embedded transistors shown here with other inorganic transistors and memristors in literature, in Table I. In addition to the dual-mode synaptic characteristics that the PtNP devices offer, they yield a very large memory window in their electrical characteristics, a feature not easily replicated in other organic[14,39,40] and inorganic transistors.

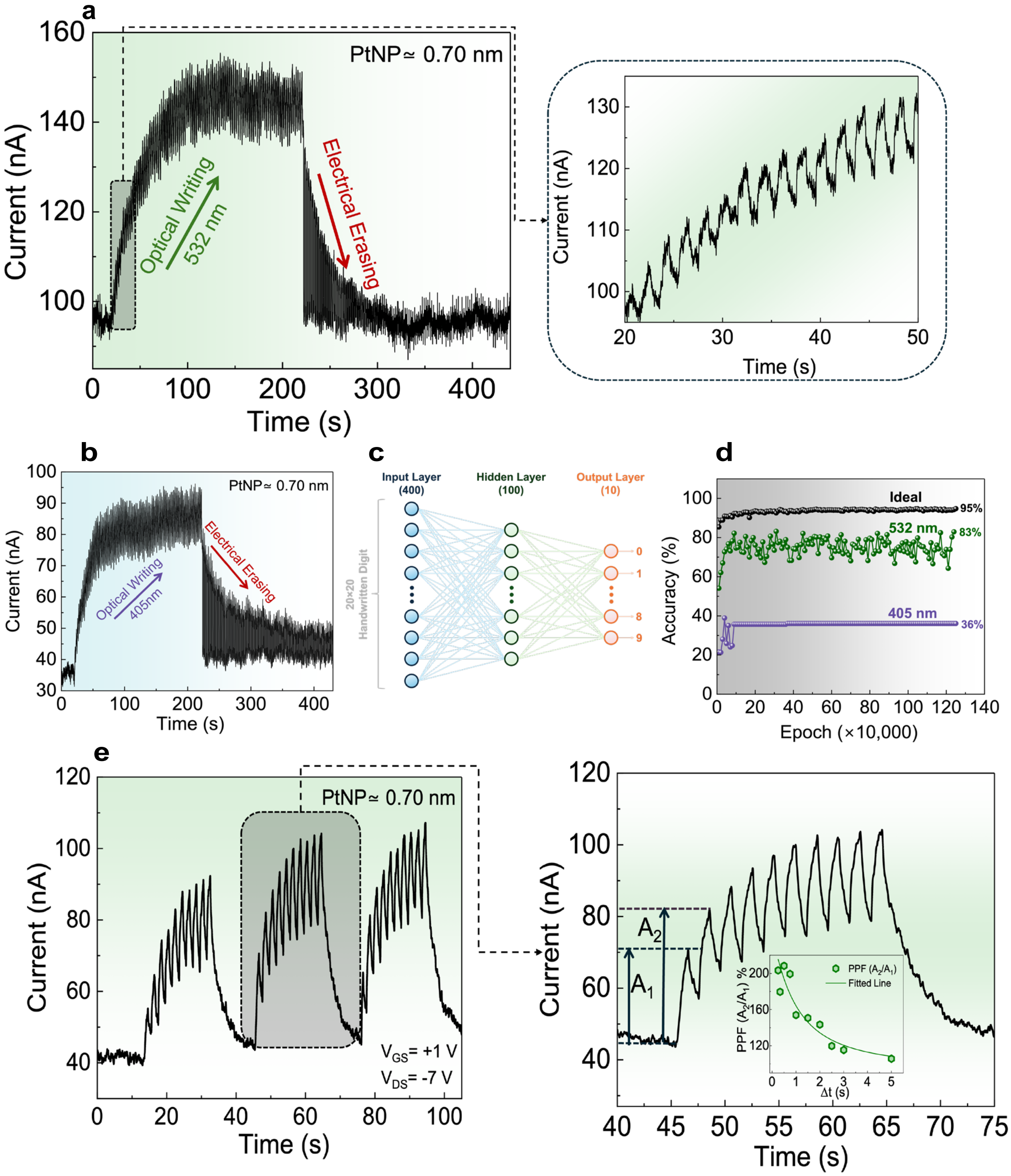

**Fig. 4| Dual programmability and neural image recognition networks from ~ 0.7 nm PtNP embedded DPP-DTT based transistors with 3 nm $Al_2O_3$ tunneling layer. a**, Current-time response under optical writing at 532 nm followed by electrical erasing. Optical writing was performed at $V_{GS} = +1$ V and $V_{DS} = -7$ V using 532 nm light pulses with a power density of 0.03 $Wcm^{-2}$ and a pulse timing of 1 s on/off. Electrical erasing was performed using $V_{GS} = +2$ V gate pulses with the same pulse timing. The zoomed-in view on the right shows the stepwise increase in current during optical programming. **b**, Current-time response under optical writing at 405 nm with a power density of 0.08 $Wcm^{-2}$, followed by electrical erasing under the same biasing and pulse timing conditions as the 532 nm. **c**, A schematic of a two-layer multilayer perceptron neural network. The network uses 400 input neuron, which corresponds to 20 × 20 pixel MNIST images, followed by 100 hidden-layer neurons and 10 output neurons. **d**, Simulated image recognition accuracy obtained using experimentally extracted parameters with the 532 nm and 405 nm illumination cycles along with a simulation for an ideal device. **e**, Current-time response under paired optical pulses (532 nm illumination)

measured at $V_{GS}$ = +1 V and $V_{DS}$ = −7 V. The zoomed-in view shows the first and second post-synaptic response amplitudes, $A_1$ and $A_2$, relative to the baseline current. The inset shows the paired-pulse facilitation response, with a maximum PPF index of approximately 210%.

**Table 1.** Comparison of dual-mode optoelectronic synaptic transistor devices.

| Device architecture | Electrical Memory window (V) | Optical condition | Control and functionality | PPF ratio | Refer-ences |
|---|---|---|---|---|---|
| DPP-DTT/PVDF-HFP organic transistor with $Al_2O_3$/PtNP/$Al_2O_3$ | 21 | 532 nm (30 mW/cm²); 405 nm (8 mW/cm²) | Dual optical (LTP)/electrical (LTD) modulation; STP-to-LTP transition | ~210% | This work |
| $ReS_2$/HZO MFMIS FeFET | 10.5 | 658 nm, 0.5 s 636.9 pW optical synaptic operation | STP-to-LTP transition and optical/electrical LTP/LTD | NR | [41] |
| $SnS_2$/HZO transistor | 2.5 | NR | STP-to-LTP electrical LTP/LTD | 160% (electrical) | [42] |
| $Ta_2NiSe_5$/$SnS_2$ heterojunction transistor | NR | PPF at 635 nm 200 ms pulse Δt = 80 ms | STP-to-LTP transition and LTP/LTD | 158% | [43] |
| $MoS_2$/Au NP floating-gate array | 2 | 520 nm | PPF; optical LTP; electrical LTD/erase | ~160% | [44] |
| $BaSnO_3$ electrolyte-gated transistor | ~2 | UV light | STP-to-LTP transition and LTP/LTD | ~130% | [45] |
| Screen-printed ZnO synaptic transistor array | ~1 | 365 nm 0.23 mW/cm² | STP and LTP in dual mode | 174% optical 177% electrical | [46] |

**Note.** NR = not reported; STP = short-term plasticity; LTP = long-term potentiation; LTD = long-term depression; PPF = paired-pulse facilitation. PPF values are listed as reported in the cited works.

## Conclusions

By embedding sub-nm PtNPs between ultrathin $Al_2O_3$ layers at the organic semiconductor (DPP-DTT) and dielectric (PVDF-HFP) interface, the transistors exhibited a large non-volatile memory window, pronounced threshold-voltage shifts, and repeatable hysteretic switching that were absent in the control devices (without PtNPs). The current-voltage transfer characteristics of the transistors, when sweeping from the accumulation to the depletion region, show remarkable features consistent with Coulomb-blockade-like transport. The width of two plateaus observed in the transfer characteristics of the transistors coincide with the IP and EA energies of the PtNP clusters. The abrupt decrease in the current, at the end of the plateau regions, upon a small voltage change reflects a change in the charge state from negatively charged PtNPs to their neutral state, followed by positively charged PtNPs. Additionally, the transistor channel current experiences an oscillating screened voltage from the electrons tunneling in and out of the PtNP clusters, which manifest as evenly spaced current spikes. A phenomenological model was developed for the oscillating current versus voltage characteristics, yielding a similar transition voltage rate across all devices. Our results demonstrate remarkable uniformity and collectivity in the tunneling behavior of the PtNPs.

Beyond electrical memory behavior, the PtNP-engineered transistors demonstrated dual optical and electrical programming, producing long-term potentiation and depression-like signals, and optically induced PPF characteristics. The optical writing with pulsed illumination and electrical erasing with a pulsed gate voltage changes the conductance states, analogous to the synaptic weight update during biological synapses. The nonlinear weight updates were benchmarked using a neural network for pattern recognition with image recognition accuracy reaching ~83% when excited with the 532 nm light. The STP is reflected with an enhanced PPF index of 210%. The extracted time constants show the coexistence of fast and slow relaxation processes, mimicking short-term biological plasticity.

The PtNP-based interfacial stack, within an organic transistor, functions as both a robust charge-storage medium for non-volatile memory and an enabling platform for optical and electrical synaptic operation. This works opens future investigations of integrating organic transistor-based memory elements with sensor interfaces, synaptic arrays, or hybrid silicon-based circuitry in platforms where each device stores a stable and tunable state, while also exploiting the individual current spikes for controlling the conductance states.

## Methods

### Materials

Poly(vinylidene fluoride-co-hexafluoropropylene) [PVDF-HFP, Mw = 455,000 g mol$^{-1}$] was purchased from Millipore Sigma. The donor-acceptor copolymer DPP-DTT was obtained from 1-Material Inc. (Dorval, Quebec, Canada). N,N-dimethylformamide (DMF) and anhydrous 1,2-dichlorobenzene (98%) were purchased from Millipore Sigma (St. Louis, MO, USA). The platinum target used for nanoparticle deposition was purchased from Kurt J. Lesker Company. Trimethylaluminum (TMA) was purchased from Strem Chemicals and used as the aluminum precursor for atomic layer deposition.

### Device fabrication

*Transistors*

1″ × 1″ glass substrates were cleaned using an organic solvent cleaning procedure. A 50 nm Al gate electrode was deposited on the cleaned glass substrates by thermal evaporation through a patterned shadow mask. The ferroelectric dielectric layer was then formed on top of the Al gate by spin-coating PVDF-HFP from solution. PVDF-HFP was dissolved in DMF at a concentration of 100 mg mL$^{-1}$, heated at 80 °C for 3 h, and stirred overnight at room temperature. To obtain thinner dielectric films, the solution was diluted to 50 mg mL$^{-1}$ prior to spin-coating. In a glove box, under a nitrogen atmosphere, the PVDF-HFP solution was statically dispensed onto the Al gate electrode and spin-coated at 1600 rpm for 60 s. The coated substrates were annealed at 70 °C for 10 min in nitrogen to remove residual solvent. This resulted in a dielectric thickness of approximately 50 nm.

Following the formation of the PVDF-HFP dielectric layer, a 12 nm $Al_2O_3$ layer was grown directly on top of the PVDF-HFP surface by atomic layer deposition (ALD) at 150 °C. Details of the ALD process of $Al_2O_3$ on PVDF-HFP are discussed in Ref. [30]. Platinum nanoparticles (PtNPs) were then deposited on the $Al_2O_3$ layer by sputtering using an AJA ATC 2000V sputtering system using the procedure discussed in Ref. [31]. The Pt target was mounted on a sputtering gun tilted at an angle of 23.8° with respect to the substrate. The deposition was carried out at room temperature. Before deposition, the sputtering chamber was evacuated to a base pressure of approximately $10^{-8}$ Torr. High-purity Ar gas with a purity of 99.999% was used as the sputtering gas, and the working pressure during PtNP deposition was maintained at 4 mTorr. Before the actual deposition, the Pt target was pre-sputtered for 10 min at an RF power of 100 W to condition the target surface and remove possible surface contamination. PtNPs were then deposited using RF sputtering at 13.56 MHz with a deposition power of 30 W. Two deposition times, 10 s and 20 s, were used to obtain PtNPs with different sizes. After deposition, the samples were removed from the sputtering chamber and handled under ambient conditions. After PtNP deposition, an ultrathin $Al_2O_3$ tunneling layer with a thickness of either 2 nm or 3 nm was deposited by ALD on top of the nanoparticles. The resulting interfacial stack consisted of 12 nm $Al_2O_3$ / PtNPs with a 2 nm or 3 nm ALD grown $Al_2O_3$ tunneling layer.

The semiconducting DPP-DTT layer was deposited on top of the completed dielectric/interfacial stack. DPP-DTT was dissolved in 1,2-dichlorobenzene at a concentration of 5 mg mL$^{-1}$. The solution was sequentially heated at 100 °C for 1 h, 130 °C for 1 h, and 145 °C for more than 12 h while stirring at 200 rpm. After cooling, the solution was stirred overnight at room temperature, filtered through a 0.45 µm PTFE filter, and reheated at 145 °C for 30 min before deposition. 75 µL of the DPP-DTT solution was dynamically spin-coated onto the device stack at 900 rpm for 60 s in a nitrogen filled glove box. To limit the semiconductor film to the channel region, Teflon tape was applied during spin-coating and removed before annealing. The devices were then annealed in an oven at 120 °C for 1 h under nitrogen.

The source and drain electrodes were finally formed on top of the DPP-DTT film by thermally evaporating a Au layer of 50 nm thickness through a patterned shadow mask. The devices had fixed channel width of 1000 µm and each substrate supported four different channel lengths of 50, 75, 100, and 125 µm.

## Characterization

*Electrical Measurements*

Current–voltage characteristics were measured at room temperature using a Keithley 4200A-SCS parameter analyzer. Optically induced current–time responses during optical writing at 405 nm and at 532 nm, followed by electrical erasing were measured both using a Zurich Instrument MFLI lock-in amplifier as well as Keithley 236 source meter. The currents measured by the MFLI were corrected with an offset and the gain factor to obtain the excitatory and inhibitory currents.

*Transmission Electron Microscope (TEM)*

The morphology and size distribution of the platinum nanoparticles were characterized using high-resolution transmission electron microscopy (HRTEM) with a Thermo Fisher Scientific Spectra 300 microscope.

## Transistor parameters

The carrier mobility in the saturation region was extracted from $\mu_{sat} = (2L/WC_i)\left(\partial\sqrt{I_D}/\partial V_{GS}\right)^2$, where $W$ and $L$ are the channel width and length, respectively. $C_i$ is the capacitance/area of the dielectric, and $V_{GS}$ is the gate-source voltage. The threshold voltage, $V_{th}$, is obtained from the drain current ($I_D$) in the saturation region: $I_D(sat) = (W/2L)\mu C_i(V_{GS} - V_{th})^2$.

## Data availability

The source data of each figure in the main text are provided with this paper. Other data that support the findings of this study are available from the corresponding author upon reasonable request.

## Supporting Information

HRTEM images, current-voltage and other transistor characteristics, current-time responses for dual mode programming, image recognition accuracies for device-to-device variation.

## Author Contributions

A. Ghobadi, S. Guha, and S. Gangopadhyay conceived the work. A. Ghobadi was involved with the fabrication of devices, conducting electrical measurements, and analyzing the data. A. Abhijeet and T. Kallaos performed the optical synaptic measurements. S. Guha helped with the analysis of the data. S. Gangopadhyay provided input on the sputtering conditions of PtNPs and device architecture. J. C. Mathai helped with the ALD growth and optimized the deposition of PtNPs. T. B. Kallaos performed the neural network pattern recognition simulations. S. C. Klue modeled the current voltage data. C. A. Ullrich provided input on the theoretical aspects of PtNP clusters. The manuscript was written by S. Guha and A. Ghobadi with contributions from all authors. All authors have given approval to the final version of the manuscript.

## Notes

The authors declare no competing financial interest.

**ACKNOWLEDGMENT**

We acknowledge the support of this work through the U.S. National Science Foundation (NSF) under Grant No. ECCS-2324839. TBK acknowledges the MizzouForward Undergraduate Research Training Grant. We thank Prof. Irene D'Amico (U. York) for valuable discussions on PtNP quantum dots. We thank Tim Pieshkov and Ahmed Jasim for helping with HRTEM measurements.

*Electronic Supplementary Information*

# Coulomb blockade-like transport and enhanced memory in organic transistors embedded with sub-nm Pt nanoparticles for neuromorphic computing

Arash Ghobadi[†], Thomas B. Kallaos[†], Abhi Abhijeet[†], Stephen C. Klue[†], Joseph C. Mathai[#], Carsten A. Ullrich[†], Shubhra Gangopadhyay[#*], and Suchismita Guha[†‡*]

[†]*Department of Physics and Astronomy, University of Missouri, Columbia, MO 65211*

[‡]*MU Materials Science and Engineering Institute, University of Missouri, Columbia, MO 65211*

[#]*Department of Electrical Engineering and Computer Science, University of Missouri, Columbia, MO 65211*

* Corresponding Author E-mail: guhas@missouri.edu

## Contents

## 1. HRTEM images

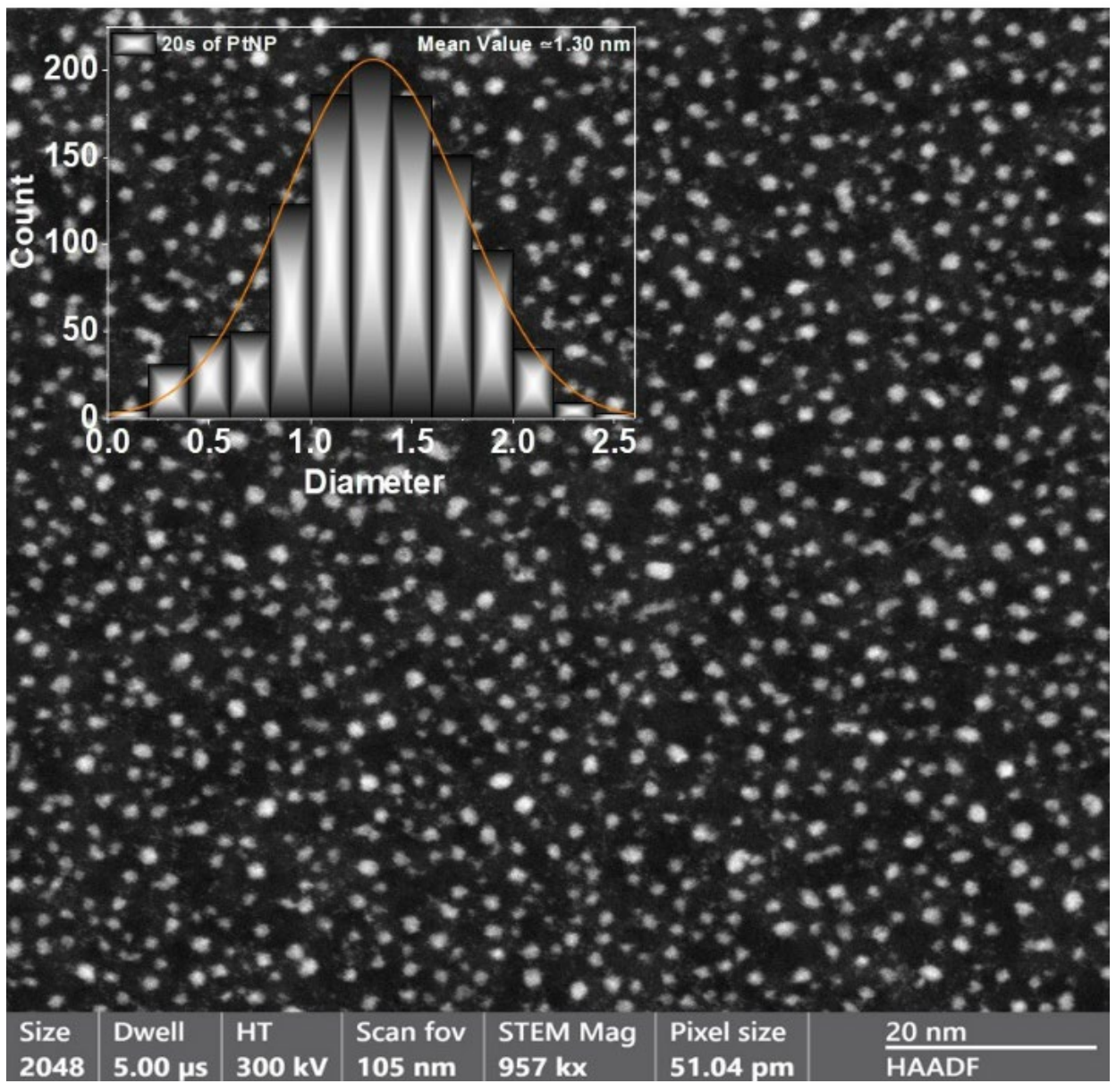

**Supplementary Fig. 1:** HRTEM image of PtNPs deposited for 20 s cycle. The inset shows a histogram of the particle size distribution with an average value of 1.30 ± 0.43 nm.

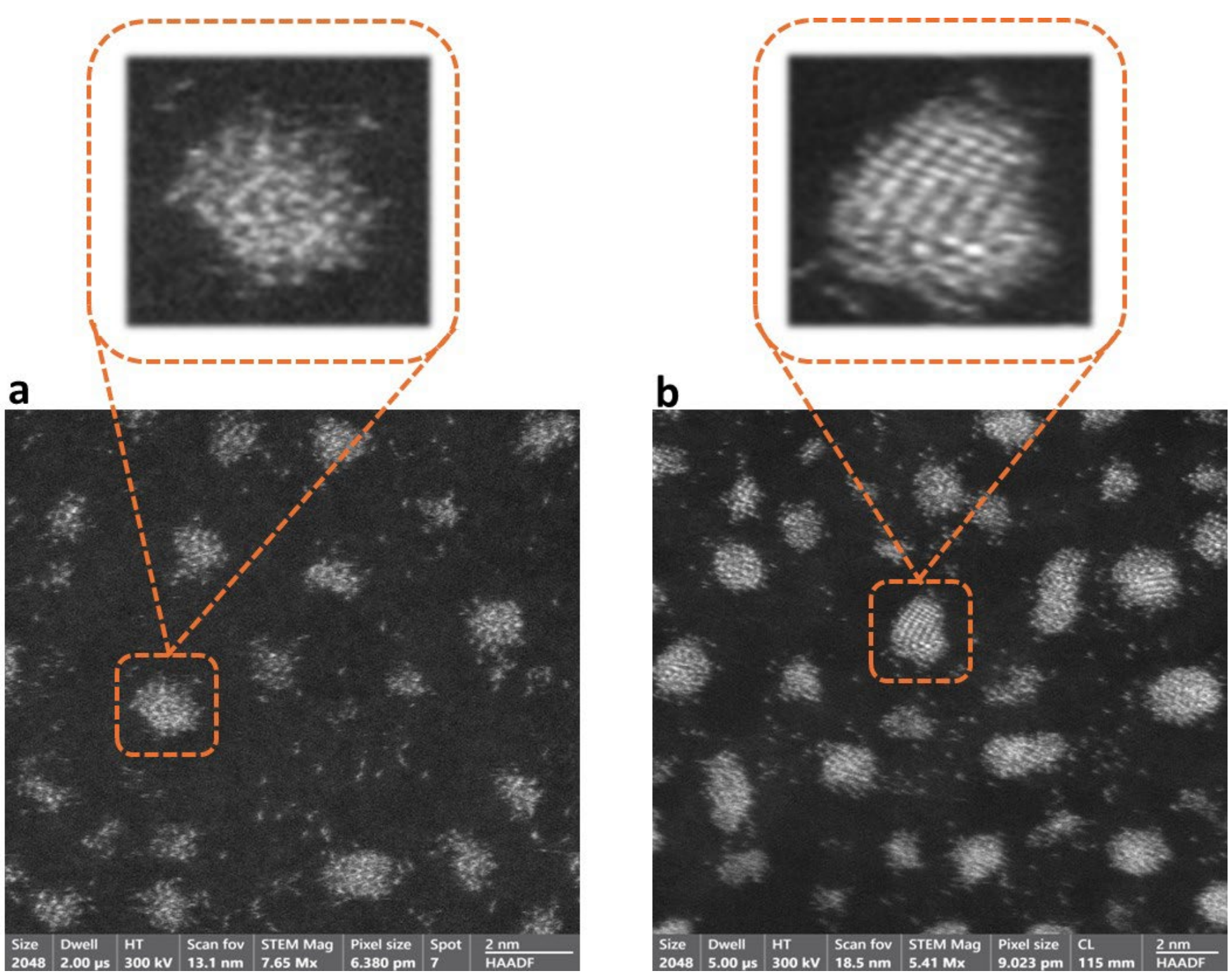

**Supplementary Fig. 2:** Comparison of HRTEM images from **a**,10 s deposition of PtNPs and **b**, 20 s deposition of PtNPs. Zoomed-in regions of the two deposition times are shown on top. The 10 s deposition time of PtNPs results in less crystalline particles compared with the 20 s deposition time.

## 2. Current-voltage and other transistor characteristics

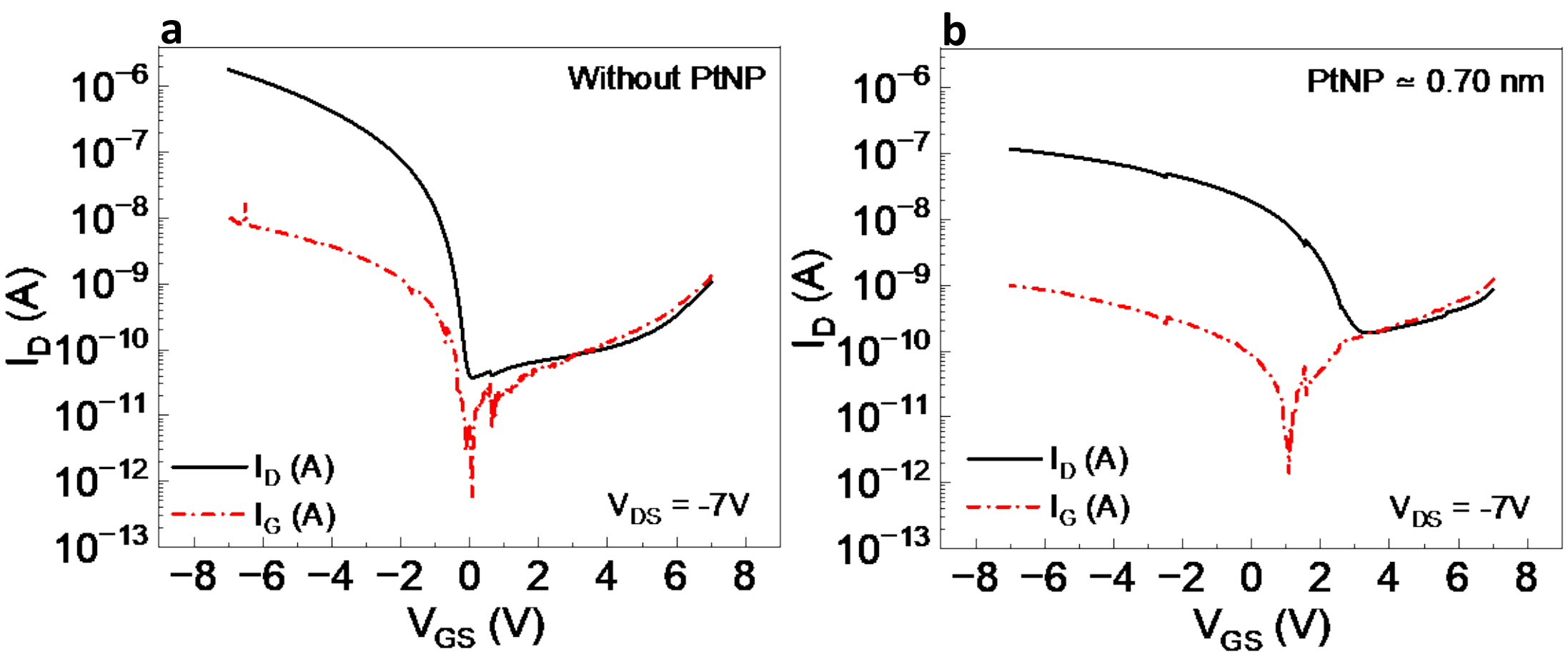

**Supplementary Fig. 3:** Transfer characteristics sweeps by varying $V_{GS}$ in **a**, control transistor without PtNPs and **b**, with 0.7 nm PtNPs and 2 nm $Al_2O_3$ tunneling layer. The black curves show the drain current, and the red curves show the gate current.

**Supplementary Table 1**. Transistor characteristics comparing devices embedded with 0.7 nm PtNPs (with two different tunneling oxide layers) and a control device. The third and fourth column list the channel width ($W$) and length ($L$). The saturation carrier mobility ($\mu_{sat}$), subthreshold swing ($SS$), threshold voltage ($V_{th}$), on on/off ratio are listed in columns five to eight.

| | Tunneling layer ($Al_2O_3$) | $W$ (µm) | $L$(µm) | $\mu_{sat}$ (cm$^{-2}$V$^{-1}$ s$^{-1}$) | $SS$ (V/dec) | $V_{th}$ (V) | *On/off* |
|---|---|---|---|---|---|---|---|
| PtNPs | 2 nm | 1000 | 75 | $1.80\times10^{-3}$ | 0.860 | 4.20 | $10^3$ |
| PtNPs | 3 nm | 1000 | 50 | $2.83\times10^{-3}$ | 0.278 | 6.62 | $10^4$ |
| No PtNPs | - | 1000 | 75 | 0.1 | 0.240 | -1.50 | $10^4$ |

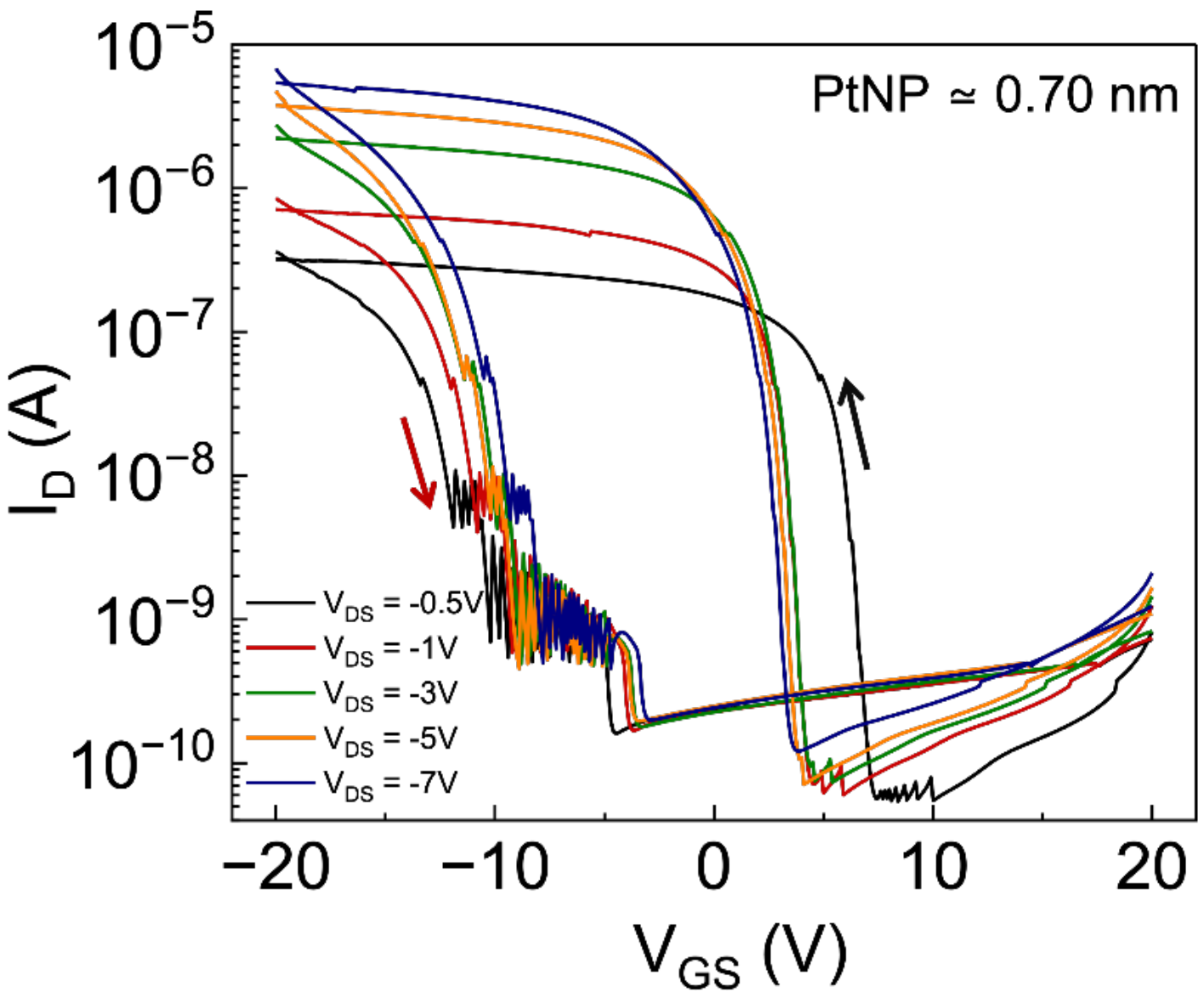

**Supplementary Fig. 4:** Transfer characteristics sweep for varying drain source voltage ($V_{DS}$) by changing $V_{GS}$ from +20 V to – 20 V and then back to -20 V for a PtNP embedded transistor with 2 nm $Al_2O_3$ as the tunneling layer. The memory window is significantly enhanced at $V_{DS}$ of -0.5 V.

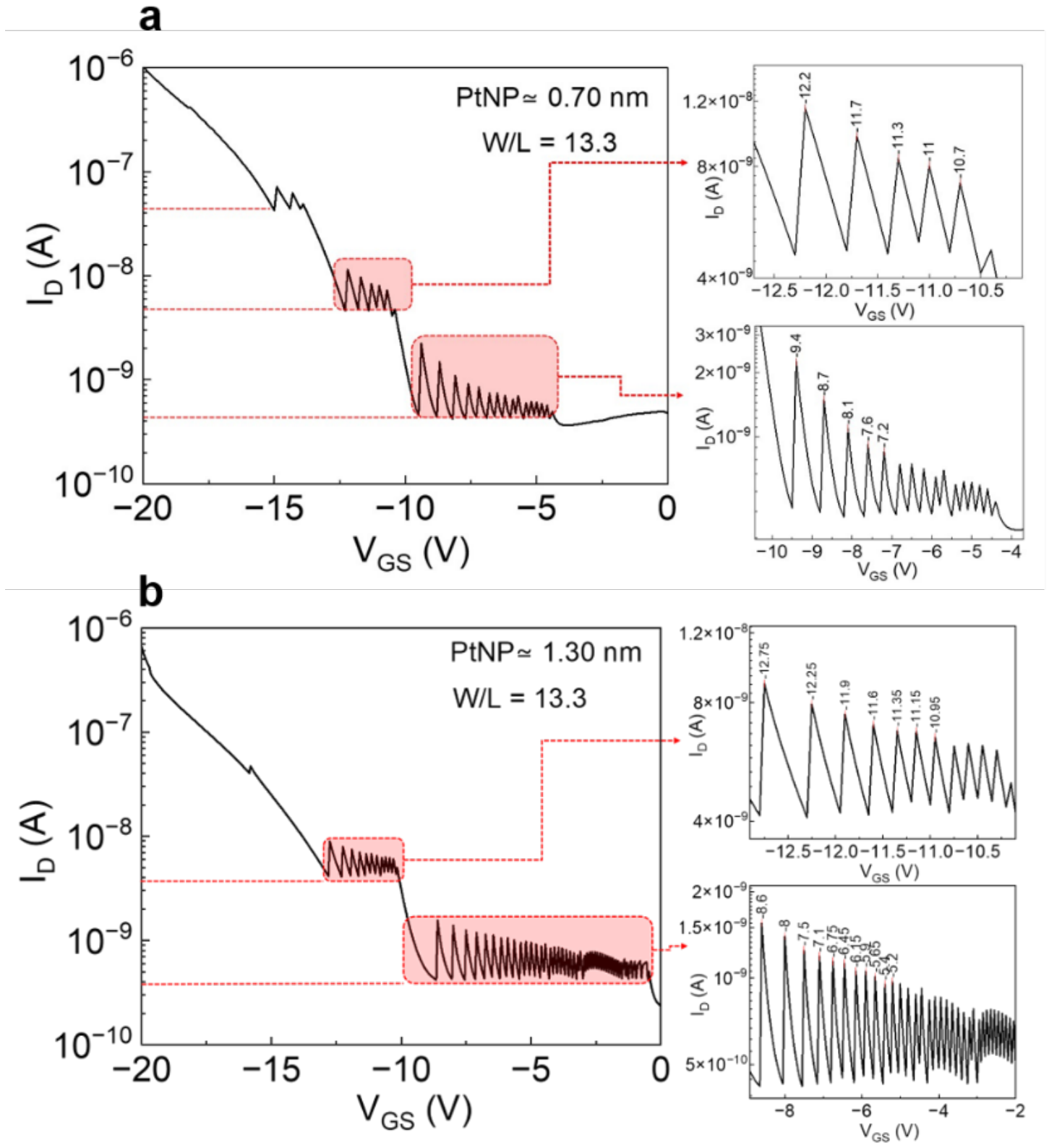

**Supplementary Fig. 5:** Transfer characteristics from accumulation to depletion for two transistor channel lengths of 75 µm. (a) Transfer characteristics of a 0.7 nm PtNP embedded transistor showing current spikes at three different plateaus. The zoomed in regions on the right show the separation between the current

spikes. (b) Transfer characteristics of a 1.3 nm PtNP embedded transistor showing current spikes at two different plateaus. The zoomed in regions on the right show the separation between the current spikes.

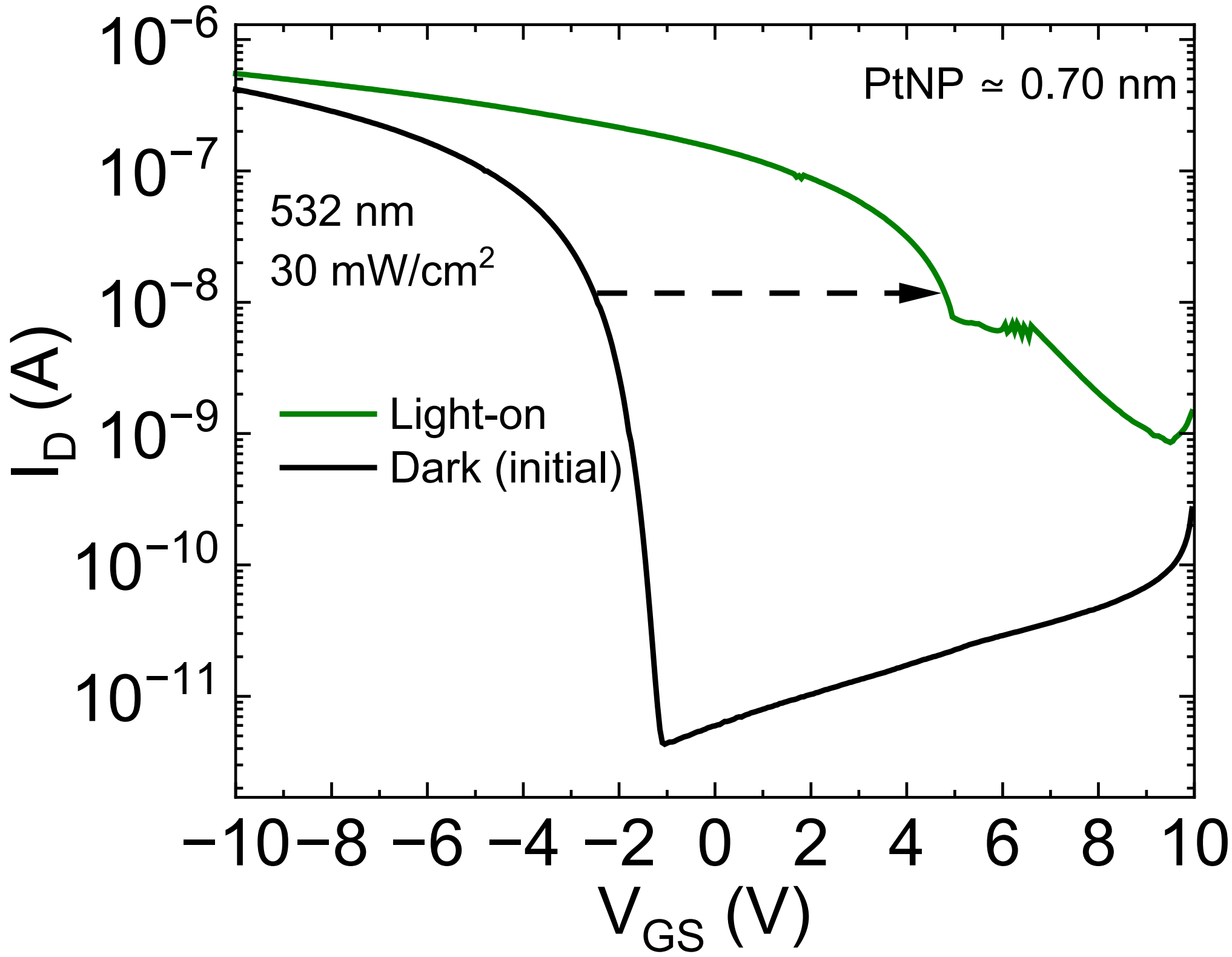

**Supplementary Fig. 6:** Transfer characteristics sweep from +10 V to 10 V under dark and 532 nm illumination for a PtNP embedded transistor with 2 nm $Al_2O_3$ as the tunneling layer.

## 3. Dual-mode programming

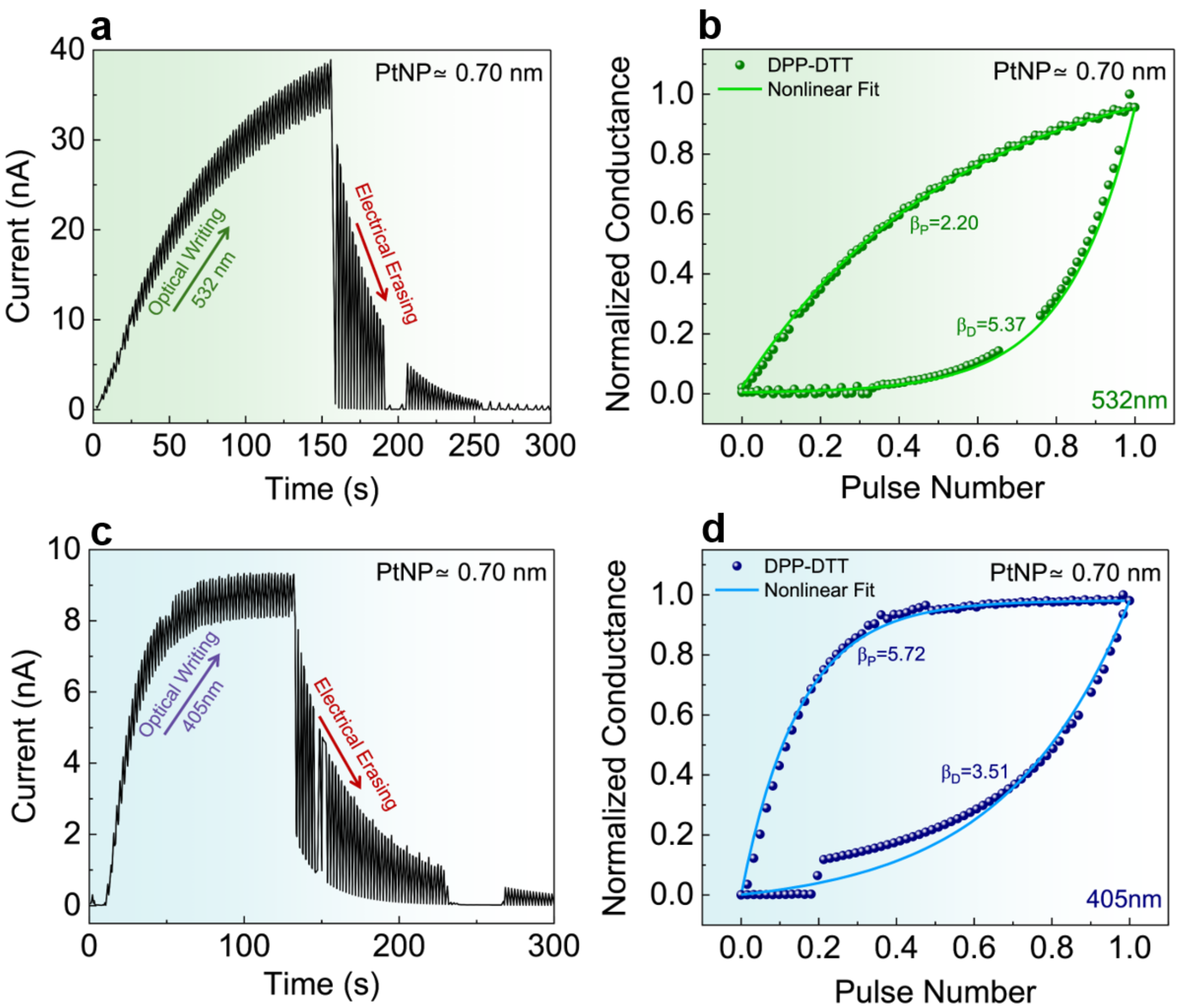

**Supplementary Fig. 7:** Current-time responses for obtaining the synaptic weight update (used for pattern recognition accuracies) under optical followed by electrical erasing. Optical writing was performed at $V_{GS}$ = +1 V and $V_{DS}$ = −7 V. (a) The current-time response with 532 nm light pulses and a power density of 0.03 Wcm$^{-2}$. (b) Normalized conductance versus normalized pulses during potentiation with the 532 nm illumination and electrical depression. The curve is fit to two nonlinear equations (bold green lines) discussed in the main section. (c) The current-time response with 405 nm light pulses and a power density of 0.008 Wcm$^{-2}$. (b) Normalized conductance versus normalized pulses during potentiation with the 405 nm illumination and electrical depression. The curve is fit to two nonlinear equations (bold blue lines) discussed in the main section.

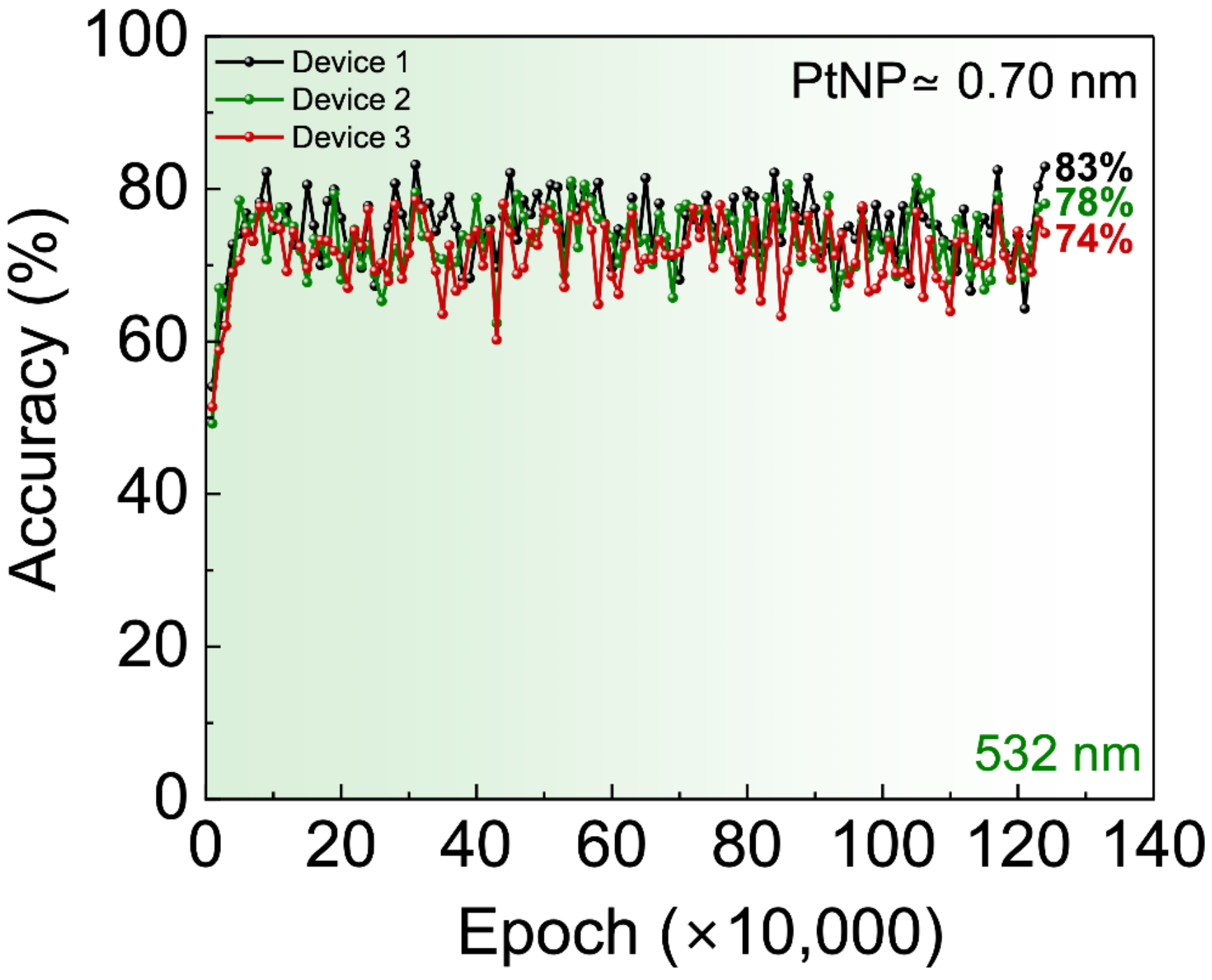

**Supplementary Fig. 8:** Device-to-device variation for simulated image recognition accuracies obtained from a 0.7 nm PtNP embedded transistor, excited with an optical illumination of 532 nm and electrical depression as discussed in the main section.